\documentclass{acm_proc_article-sp}

\usepackage{url}
\usepackage{booktabs}
\usepackage{hhline}
 \usepackage{flushend}   
 \usepackage{subcaption}
\usepackage{color}
\usepackage{array}
\usepackage{amsmath}
\usepackage[utf8]{inputenc}
\usepackage{tikz}
\usepackage{ upgreek }
\usepackage{multirow}

\DeclareMathOperator*{\argmax}{arg\,max}

\newcommand*\circled[1]{\tikz[baseline=(char.base)]{
            \node[shape=circle,draw,inner sep=1pt,color=red] (char) {#1};}}

 \title{Understanding Graph Structure of Wikipedia \\ for Query Expansion}
\numberofauthors{2} \author{ 
	\alignauthor Joan Guisado-Gámez\\ 
		\affaddr{DAMA-UPC}\\ 
		\affaddr{Universitat Polit\`ecnica de Catalunya}\\
	    \email{joan@ac.upc.edu} 
	\alignauthor Arnau Prat-Pérez\\ 
		\affaddr{DAMA-UPC}\\ 
		\affaddr{Universitat Polit\`ecnica de Catalunya}\\
	    \email{aprat@ac.upc.edu} 
}   

\newfont{\mycrnotice}{ptmr8t at 7pt}
\newfont{\myconfname}{ptmri8t at 7pt}
\let\crnotice\mycrnotice%
\let\confname\myconfname%

\begin{document}


\maketitle

\begin{abstract} Knowledge bases are very good sources for knowledge
extraction, the ability to create knowledge from structured and unstructured
sources and use it to improve automatic processes as query expansion. However,
extracting knowledge from unstructured sources is still an open
challenge~\cite{SuchanekW13}. In this respect, understanding the structure of
knowledge bases can provide significant benefits for the effectiveness of such
purpose. In particular, Wikipedia has become a very popular knowledge base in
the last years because it is a general encyclopedia that has a large amount of
information and thus, covers a large amount of different topics.  In this
piece of work, we analyze how articles and categories of Wikipedia relate to
each  other and how these relationships can support a query expansion
technique. In particular, we show that the structures in the form of dense
cycles with a minimum amount of categories tend to identify the most relevant
information.  
\end{abstract}

\section{Introduction} 

Knowledge bases such as Wikipedia, Yago or Wordnet are used in many
applications as a source of
knowledge~\cite{Arguello2008,moussa2011qasyo,Voorhees94}.  One of these
applications is query expansion, which is the process of expanding a  query
issued by a user, introducing new terms, called expansion features, in order
to improve the quality of the retrieved results. Query expansion is motivated
by the assumption that the query introduced by the user is not the best to
express its real intention. For example, \textit{vocabulary mismatch} between
queries and documents is one of the main causes of a poor precision in
information retrieval systems~\cite{metzler2007similarity}. Poor results also
arise from the \textit{topic inexperience} of the users. Users searching for
information are often not familiar with the vocabulary of the topic in which
they search, and hence, they may not use the most effective keywords. This
leads to the loss of important results due to the lack of precision when
choosing the query terms.  Thus, the challenge is to properly select the best
expansion features (terms added to the original query), that improve the most
the quality of the results. Note also that a bad choice of expansion
features may be counterproductive.

In this paper we focus on Wikipedia as a knowledge base for query expansion.
Wikipedia is a popular encyclopedia which contains a large amount of rich data
and thus,  its topic coverage is extremely broad.  The structure of Wikipedia
has been used for query expansion in different ways.
In~\cite{AlmasriBC14,Arguello2008, Egozi2011} the authors describe different
information extraction strategies by using the individual links of each
Wikipedia article, without going deeper into further relationships. In~\cite
{Guisado-Gamez2014} the authors borrow a social network community detection
metric~\cite{PerezDBL12} to extract better expansion features from Wikipedia,
assuming that a structure as simple as a transitive relation is sufficient to
capture good relationships among terms. However, they do not take into account
the difference between a social network and a knowledge base.

To the best of our knowledge, this work presents the first analysis of the
trends that appear in the structure of Wikipe-dia that contribute to identify
the best expansion features for a given query.  For this task, we support our
analysis on an information retrieval benchmark borrowed from the ImageCLEF 2011
track~\cite{tsikrikaPK11}, which consists of a set of documents and a set of
queries. For each query, it also contains the set of documents that are correct
results for that particular query, which from now on we will refer to as the
result set.  We use this information to build a ground truth that relates each
query from the query set to a graph of Wikipedia articles and categories that
we called query graph.  Given a query, its query graph contains those articles
that in case of being used as expansion features, allow us to retrieve the
correct documents for that particular query. From the analysis of the structure
of the query graphs we reveal that, within the maze of relations among articles
and categories that group them, cycle-based structures contribute to find
articles whose titles are good candidates to be used as expansion features.


The main contributions of this paper can be summarized as follows:
\begin{enumerate}     \item  We create a ground truth consisting      
of those articles in Wikipedia that provide good results for each of the
queries of ImageCLEF 2011 track, which we use as the baseline in our
experiments\footnote{\scriptsize The ground truth is available in:\\
https://github.com/DAMA-UPC/QueryGraphs}. 

    \item  We analyze how the articles and categories of the ground truth are
      structured within the Wikipedia graph.

    \item  We identify cycles of articles
      and categories as an important structure and also, we identify some trends 
      within them. We find that dense cycles with a minimum ratio of categories, around the 30\%,
      are able to identify the best expansion features.

    \item We identify challenging and open problems for graph processing
      technologies when it comes to exploit structures of large graphs such as
      Wikipedia. 
 \end{enumerate}  

The remainder of this paper is organized as follows: in
Section~\ref{studyDataset} we describe in detail the process of building a
ground truth out of ImageCLEF 2011. In Section~\ref{sec:analysis} we analyze
the query graphs and finally, in
Section~\ref{challenge} we  conclude and propose some challenges for graph-
based technologies.

\section{Building the ground truth}\label{studyDataset}
Query expansion consists in reformulating an input query to improve its
retrieval performance. The input query is expressed as a list of
keywords, for example, \texttt{"Graffiti Street Art"} has 3 keywords. Query
expansion combines the original query keywords with a set of expansion features that
are identified by applying morphological transformations or by finding
synonyms and semantical related concepts.

\begin{figure}
\centering
\includegraphics[width=.5\linewidth]{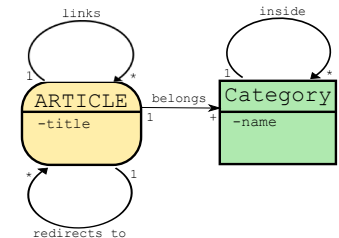}
\caption{Wikipedia Diagram.}
\label{fig:wikipedia}
\end{figure}

To find the expansion features, we rely on Wikipedia.
Wikipedia has a rich schema and can be used as a knowledge base in several
ways. In this paper we use that part of the schema depicted in
Figure~\ref{fig:wikipedia}, which consists of two different types of entries:
\texttt{Article} and \texttt{Category}.

A Wikipedia article describes a single topic, and has a title that,
according to the Wikipedia edition rules, must be \textbf{recognizable},
\textbf{natural}, \textbf{precise}, \textbf{concise} and \textbf{consistent}.
Each article represents an entity -- something that exists in itself, actually
or potentially, concretely or abstractly, physically or not --. Hence, titles
are useful to identify the entities that are mentioned in the input query. In
the example above, we identify 2 entities: \texttt{"Graffiti"} and
\texttt{"Street Art"}.

Articles can \texttt{link} to other articles and must belong to, at least, one
\texttt{Category}. Articles can also be connected by another special
kind of relation, called \texttt{redirect}, when two articles refer to the same
topic but have different titles.  In this case, the articles with the less
used/common titles (\textit{redirect articles}) points to the article with the
most common title (\textit{main article}).  Each category can also be
\texttt{inside} one or more general categories forming, according to Wikipedia
edition rules, a tree-like structure. This forms a graph with multiple nodes,
articles and categories, and relations with semantics such as equivalence,
hierarchical or associative.

\begin{table}
  \begin{tabular}{>{\centering}m{1.25cm}||m{6.75cm}}
    \hline
    \hline
    Notation&Definition\\
    \hline
    $D$&$D=\{d_1,\cdots,d_{|D|}\}$ is a set of documents.\\
    $A$&$A=\{a_1,\cdots,a_{|A|}\}$ is a set of articles.\\
    $c$&Wikipedia category.\\
    $k$&A list of keywords.\\
    $q$&A tuple $<k,D>$ such that $\forall d \in D$, $d$ is a correct document for $k$.\\
    $\mathcal{L}(k)$&The set of Wikipedia articles mentioned in $k$.\\
    $\mathcal{L}(d)$&The set of Wikipedia articles mentioned within the text of document $d$.\\
    $\mathcal{L}(D)$&$\bigcup_{d \in D}\mathcal{L}(d)$.\\
    $\mathcal{X}(q)$&The set of Wikipedia articles whose titles are the best expansion features for query $q$.\\
    $G(q)$&The query graph of $q$.\\
    \hline
\end{tabular}
\caption{Table of definitions.}
\label{tab:definitions}
\end{table}

We are interested in knowing whether the graph structure of the articles and
their categories encodes information that could potentially be used to identify
expansion features. For that purpose, we need a ground truth that relates a
query to a graph of articles and categories, which we call the query graph.
The articles of the query graph are those whose titles  a) identify the
entities mentioned in the query and b) are the best expansion features for the query. The
categories of a query graph are the categories of the articles and help to
understand better the structures. We first describe the process for building up
the ground truth based on query graphs and later, we analyze their structure in
order to identify trends that benefit the identification of good
expansion features. Before continuing, in Table~\ref{tab:definitions} we
introduce some definitions and notation that will be used throughout this
paper.

To build the ground truth we rely on the ImageCLEF 2011
track~\cite{tsikrikaPK11}. It consists of a set of 237,434 images and their
respective XML metadata files. Also, the track provides a set of fifty queries
consisting of a set of keywords $k$ and their results set $D$, as explained in
Table~\ref{tab:definitions}. These results sets contain only the XML metadata
files describing the images.
For each query $q$ of the ImageCLEF 2011 query set, we build
a query graph $G(q)$, whose construction can be summarized as follows:

\begin{enumerate}
  \item Identify the sets of Wikipedia articles $\mathcal{L}(q.k)$ and
    $\mathcal{L}(q.D)$.
  \item Find the set $\mathcal{X}(q)$.
  \item Assemble the query graph $G(q)$.
\end{enumerate}

\subsection{Linking with Wikipedia}\label{sec:entityLinkage}

\begin{figure}[t]
\centering
\includegraphics[width=.8\linewidth]{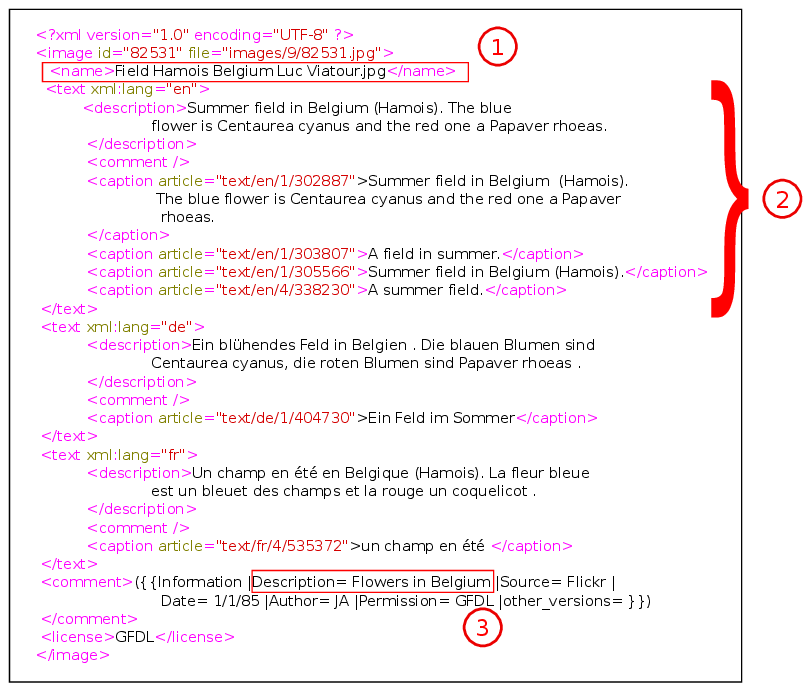}
\caption{ImageCLEF XML file.}
\label{fig:xmlFile}
\end{figure}

Given a query $q$, to identify those Wikipedia articles that are mentioned
within $q.k$ and $q.D$, we perform an entity linking process consisting in
identifying the entities within the given text. As shown in
Table~\ref{tab:definitions}, this process is denoted as $\mathcal{L}$. Even
though the entity linking process is essentially the same regardless of
whether the input item is a set of keywords or a document, for the later an
additional preprocessing step is performed where, the relevant text of the
document to be linked is extracted. Given $d$, an XML metadata document
as the one depicted in Figure~\ref{fig:xmlFile}, we extract \circled{1} the
name of the file without the file extension, \circled{2} the information in
the English section (there are also sections in German and French) and
\circled{3} the description from the general comment field. These three items
are then combined in a string, in which we do entity linking.

To perform the entity linking process, we require a knowledge base of entities
such as Wikipedia. In our case, we consider each article in Wikipedia as an
entity, whose title is used to perform the entity linking process against the input
text.  Thus, this allows us to represent a given text as a set of articles of
Wikipedia.  The entity linking process consists in identifying the set of the
largest substrings in the input query that matches with the title of an article
in Wikipedia. In order to improve the accuracy of our entity linkage, we do not
only search entities in the input text, but also in synonym phrases. We derive
a synonym phrase by replacing at least one term of the input text by a
synonymous term.  Synonymous terms are calculated using redirections of
Wikipedia. With more detail, given a term $t$, we retrieve (if it exists) the
article $a$ from Wikipedia whose title is equal to $t$. Then, the synonyms of
$t$ are the titles of the redirects of $a$. This simple strategy proved
effective for our purposes. Finally, for each query $q$ we compute
$\mathcal{L}(q.k)$ and $\mathcal{L}(q.D)$.

\subsection{Finding the best expansions $\boldsymbol{\mathcal{X}(q)}$}

According to Table~\ref{tab:definitions}, $\mathcal{X}(q)$ is the set of
articles whose titles are the best expansion features for $q$. To find
$\mathcal{X}(q)$, we need a mechanism to evaluate how good are the titles of a
set of articles $A$ when these are used as expansion features of a query $q$.
For that purpose we rely on the INDRI search engine~\cite{strohman2005indri}.
Given the articles in $A$, we use their titles to internally write a query in
the INDRI query language, based on exact phrase matching.  The returned results
are then used to calculate the top-$r$ precision of the query. So, if
$\mathcal{T}(A,r)$ is the top-$r$ results when the titles of articles in $A$
are used to write the query, then the top-$r$ precision over a set of expected
result $D$ is computed as follows:
{
\small \begin{equation*}   \mathcal{P}(A,r,D) =
\frac{|\mathcal{T}(A,r) \cap D|}{r},  \end{equation*}
}then, the average of the top-1, top-5, top-10 and top-15 precision is computed as:
{\small
\begin{equation}\label{eq:o}
  \mathcal{O}(A,D) = \frac{\sum_{r \in \mathcal{R}} \mathcal{P}(A,r,D)}{|\mathcal{R}|}, 
\end{equation}
}where $\mathcal{R}=\{1,5,10,15\}$. Note that $\mathcal{L}(q.k)$ and $\mathcal{L}(q.D)$
are the sets of articles that are mentioned in the query keywords ($q.k$) and in 
the documents of the query result set ($q.D$) respectively. Since we want to analyze 
how the articles in $\mathcal{L}(q.D)$ help to improve the most the results obtained
by $\mathcal{L}(q.k)$, we define $\mathcal{X}(q)$ as:
%
%
{\small
\begin{equation*} \mathcal{X}(q) = \argmax_{A' \subset
\mathcal{L}(q.D)} \mathcal{O}(\{\mathcal{L}(q.k) \cup A'\},q.D) 
\end{equation*}
}

The naive way to compute $\mathcal{X}(q)$ is to compute the quality for all
possible combinations of $A'$ from articles in $\mathcal{L}(q.D)$.  However, the number
of possible combinations is
{\small\[\sum_{i=1}^{|\mathcal{L}(q.D)|}{|\mathcal{L}(q.D)| \choose
i},\]}which makes unfeasible to find the best solution using a brute force
approach. Therefore we propose the following procedure to find the best
combination.

The procedure starts with $A'$ containing a random article of
$\mathcal{L}(q.D)$. From this moment on, it starts an iterative process that
incrementally applies a single operation out of the following possible:
\texttt{ADD} a new article to $A'$ from $\mathcal{L}(q.D)$,
\texttt{REMOVE} an article from $A'$, \texttt{SWAP} an article of
$A'$ by one of $\mathcal{L}(q.D)$.  
 Operations are applied
as long as they improve Equation~\ref{eq:o}, repeating the
process until no further improvements can be found. Note that if after
removing an article the quality remains the same, the article is removed as we
want the minimum set of articles with the maximum quality.


This method for building $\mathcal{X}(q)$ as $\mathcal{L}(q.k) \cup A'$ is 
capable of achieving good results in
terms of precision for the different top-$r
$. Table~\ref{tab:precision_stats}
shows, for each top-$r$, the average of the precision obtained in all the queries
for that particular \mbox{top-$r$}.
\begin{table}[t]
  \centering
  \begin{tabular}{l||r r r r r }
    \hline\hline 
     &\multirow{2}{*}{min}&\multicolumn{3}{c}{Quartiles}&\multirow{2}{*}{max}\\
     &  & 25\% & 50\% & 75\% & \\
    \hline
    top-1&0&1&1&1&1 \\
    top-5&0&1&1&1&1\\
    top-10&0.2&0.6&0.9&1&1\\
    top-15&0.2&0.65&0.8&0.85&1\\
    \hline
  \end{tabular}
  \caption{Statistics of precision of ground truth.}
  \label{tab:precision_stats}
\end{table}

\begin{figure}
\centering
\includegraphics[width=\linewidth]{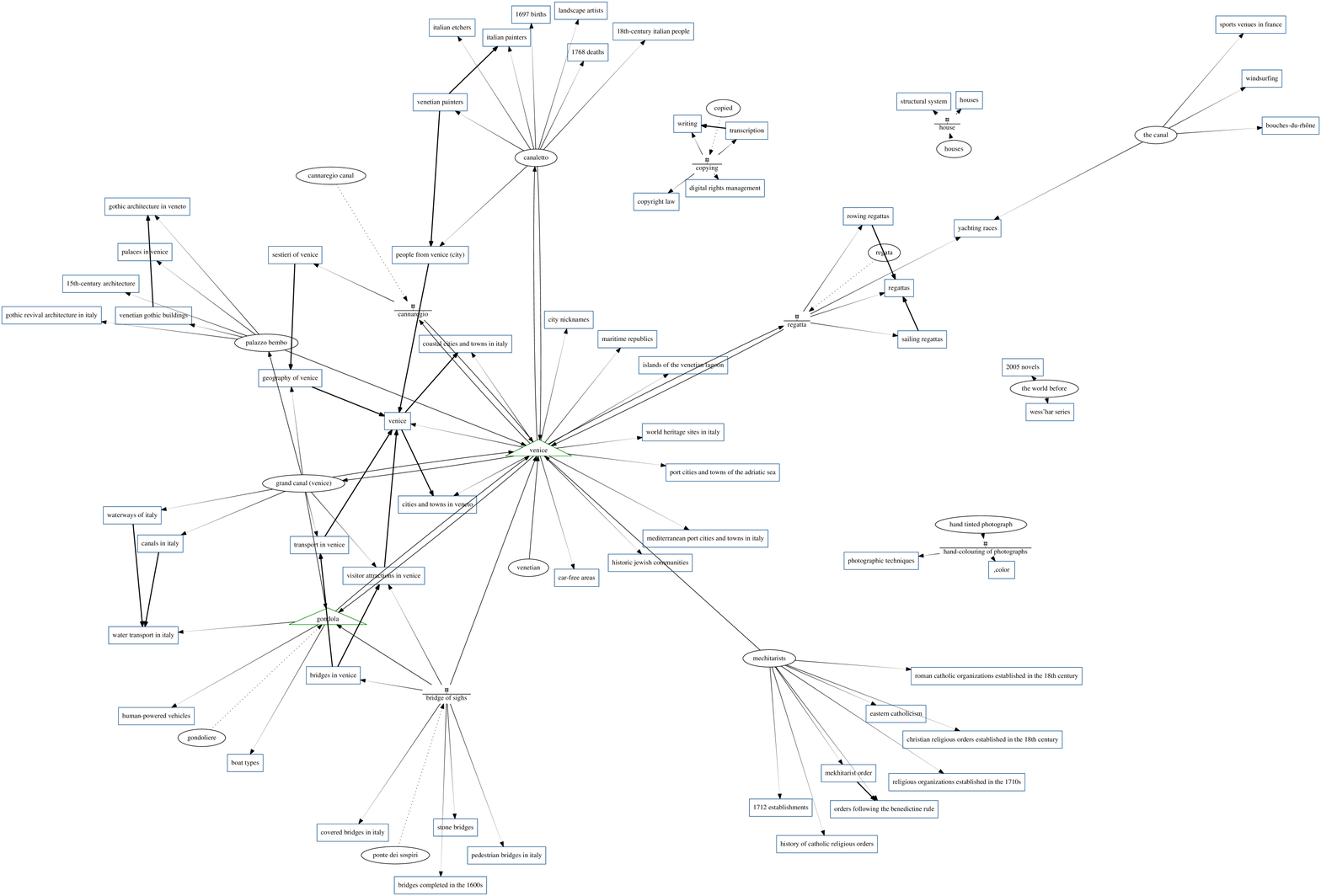}
\caption{Overview of the graph query of query \#90  ``gondola in venice''.}
\label{fig:queryGraph}
\end{figure}

\subsection{Query Graph Assembly}

Finally, each query graph $G(q)$ is built by inducing the subgraph with nodes
$\mathcal{X}(q)$, their main articles in case of being a
redirect (see Section~\ref{fig:wikipedia}), and their categories. This allows
us to build $G(q)$ as a representation of the entities in the query, the
expansion features that contributes the most in terms of precision, and also
the semantics  provided by the categories, making $G(q)$ a good representation
of the query domain.  Figure~\ref{fig:queryGraph} depicts an example of $G(q)$
for one of the queries of ImageCLEF. Articles that belong to
$\mathcal{L}(q.k)$ are depicted with a
triangular box, articles in $A'$ in circles boxes, squared boxes are used
for categories, and finally, the nodes that do not have a box are the main
articles.

\section{Query graph analysis}\label{sec:analysis}

A quick analysis of the query graphs reveals that they are, in general,
disconnected graphs composed by a moderately large connected component. This
is observed in Table~\ref{tab:cc_stats}, where we show the minimum, the first,
second and third quartiles and the maximum of the relative size, relative number of
query nodes, ratio of articles and categories and the expansion ratio, which
is $|\mathcal{X}(q)|/|\mathcal{L}(q.k)|$, of the largest connected component of
all query graphs.

\begin{table}
  \begin{tabular}{l||r r r r r }
    \hline\hline 
    &\multirow{2}{*}{min}&\multicolumn{3}{c}{Quartiles}&\multirow{2}{*}{max}\\
     &  & 25\% & 50\% & 75\% &  \\
    \hline
    \%size&0.164&0.477&0.587&0.688&1 \\
    \%query nodes &0&1&1&1&1\\
    \%articles&0.025&0.148&0.217&0.269&0.5\\
    \%categories&0.5&0.731&0.783&0.852&0.975\\
    expansion ratio&0&2.125&4.5&23.750&176\\
    \hline
  \end{tabular}
  \caption{Statistics of the largest connected component of the query graphs.}
  \label{tab:cc_stats}
\end{table}

We see that this large connected component contains, in general, all articles
of $\mathcal{L}(q.k)$.  This is an interesting observation as it means that,
in general, the terms users introduce in a search engine are semantically
related either directly or by means of extra articles or categories. This
suggests that Wikipedia, as we will see in more detail shortly, contains this
semantic relation encoded within its structure, and therefore, can be
exploited.  Also, we observe that the largest connected component is clearly
dominated by categories (makes sense, since each article belongs, at least, to
one category), and that the number of expansion features introduced per
article in $\mathcal{L}(q.k)$ goes from 0 (which we use to denote that no
article of $\mathcal{L}(q.k)$ was in this connected component), to 176. This last
result suggests that the variety of queries of the benchmark is large, ranging
from queries whose graphs touch a very local region of Wikipedia, to others
where very distant terms are connected.

We also detected that, compared to the other connected components, which
consist of a single article and its categories and thus, its structure is not
worth to be analyzed in detail, the largest connected component is
significantly structured. The average \textit{triangle participation ratio
(TPR)} of the largest connected components is around \textbf{0.3}. TPR counts
the ratio of nodes that belong to at least one triangle. This value is
particularly large if we consider that the category graph in Wikipedia is
tree-like and therefore triangles are not present.  Furthermore, besides the
triangles, we also observe a significant presence of cycles of lengths 
two, four and five. Figure~\ref{fig:queryGraph} shows the query graph of query
90 ``gondola in venice''. We see that there is a large connected component
containing most of the nodes, more structured than the other four smaller
connected components. In this example, the expansion ratio is about 6.5, with
expansion features being up to distance three from query articles.

Since there is a considerably large region of these query graphs with articles
and categories structurally related, we focus on analyzing the way they are
interconnected. More concretely, we analyze the cycles they form, and how their
characteristics correlate with the quality of the expansion features their
contain. We define a cycle $C$ as a sequence of $|C|$ nodes (either articles or
categories) starting and ending at the same node, with at least one edge among
each pair of consecutive nodes.  $|C|$ denotes the length of the cycle. Note
that this description allows a cycle $C$ to contain a subcycle $C'\subset C$
(cycle within a cyle) of length $|C'| < |C|$, as we do not enforce the cycles
to be cordless. In our definition,  we do not consider the direction of the
edges, and we limit the length of the cycles to 5 as the cost of finding the
cycles grows exponentially with the length of the cycle.  Finally, we are
interested in those cycles containing at least one article of
$\mathcal{L}(q.k)$, as we want to know how other articles and categories relate
to the original articles of the query.

\begin{figure}[!t]
\begin{centering}
    \begin{subfigure}[b]{0.2\linewidth}
      \begin{centering}
      \includegraphics[width=\linewidth]{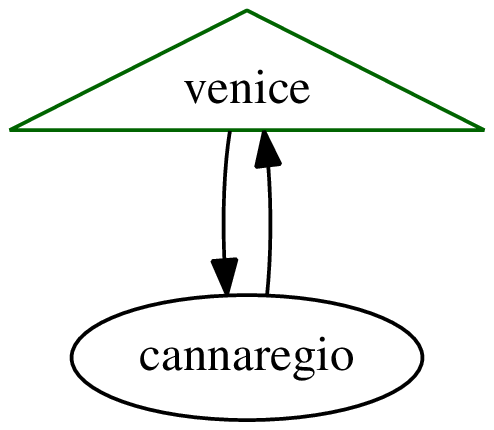}
      \caption{}
      \label{subfig:cycle2}
      \end{centering}
    \end{subfigure}
    \begin{subfigure}[b]{0.38\linewidth}
\begin{centering}
      \includegraphics[width=\linewidth]{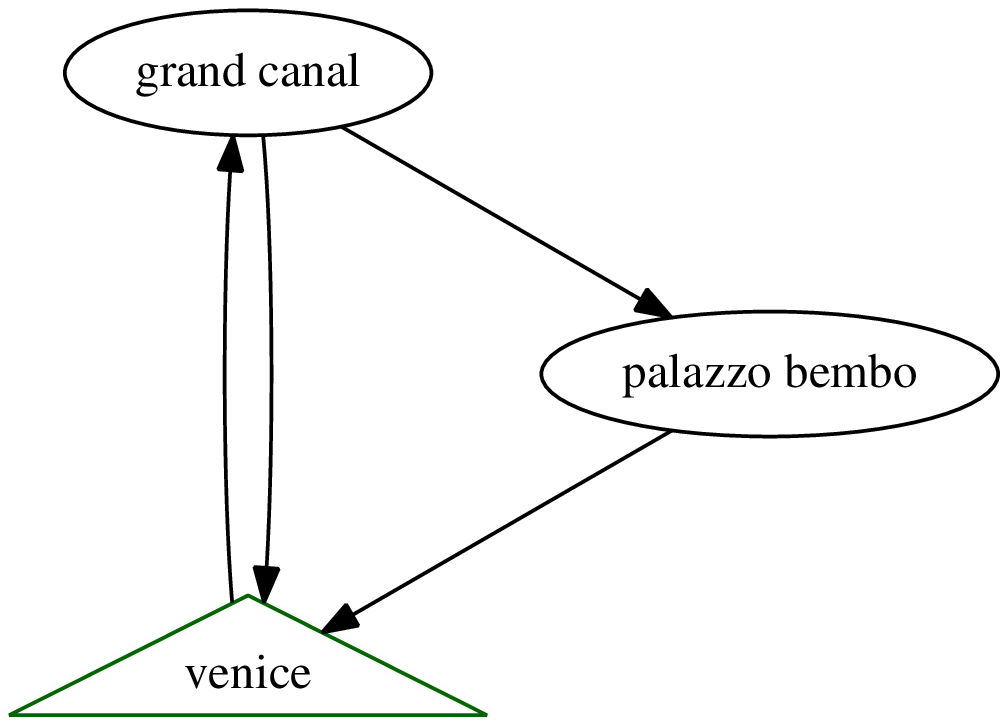}
      \caption{}
      \label{subfig:cycle3}
      \end{centering}
    \end{subfigure}
    \begin{subfigure}[b]{0.39\linewidth}
\begin{centering}
      \includegraphics[width=\linewidth]{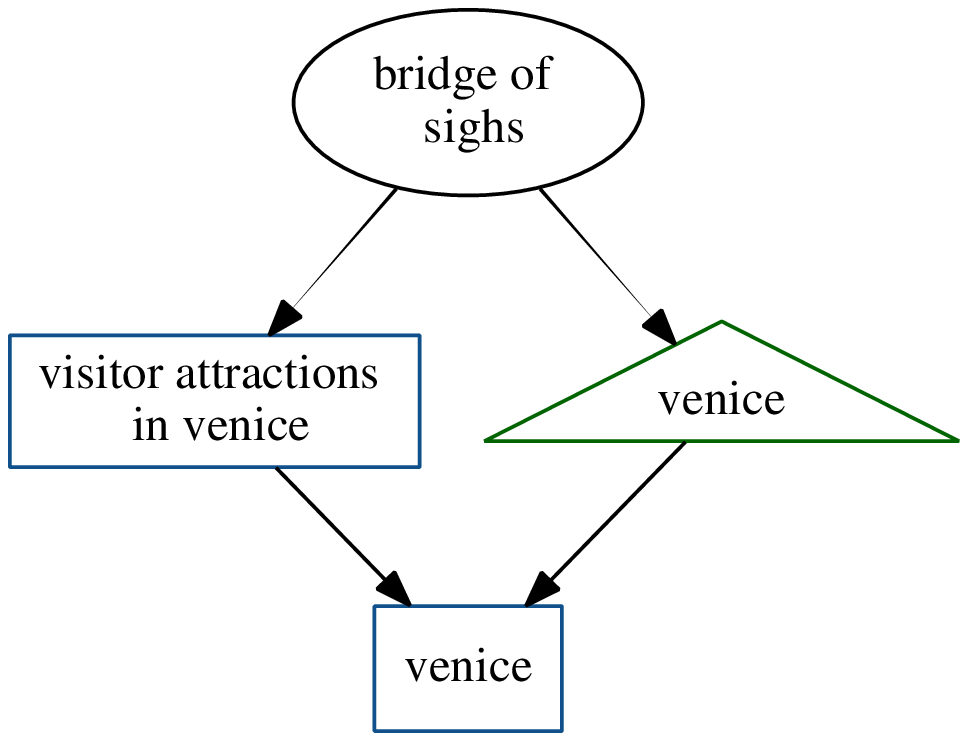}
      \caption{}
      \label{subfig:cycle5}
      \end{centering}
    \end{subfigure}
    \caption{Cycles of length 2~(a), 3~(b) and 4(c).}\label{fig:cycleTypes}
    \end{centering}
\end{figure}

Following the example of query 90 ``gondola in venice'', in
Figure~\ref{fig:cycleTypes} we show three examples of cycles of lengths 2, 3
and 4, that due to the relations established between their articles and
categories, are capable of linking semantically related concepts.  For
example, in Figure~\ref{fig:cycleTypes}(a) a cycle of length 2 introduces the
expansion feature \texttt{"cannaregio"} -- the northernmost of the six
historic districts of Venice, whose main artery is the Cannaregio Canal, a
gondola navigable canal. In Figure~\ref{fig:cycleTypes}(b) due to a cycle of
length 3, the expansion features \texttt{"grand canal"} and \texttt{"palazzo
bembo"} are introduced, and \texttt{"bridge of sighs"} is an expansion feature
introduced by a cycle of length 4, as shown in Figure~\ref{fig:cycleTypes}(c).
All these expansion features are popular attractions in Venice and likely to
be surrounded by gondolas\footnote{\scriptsize A quick search of these expansion features
in Google Images confirms this statement.}. These observations regarding
the properties of cycles are worthy of further analysis.

\begin{table}[ht]
\centering 
\begin{tabular}{r|| c c c c c} 
\hline\hline 
Cycle Size  & Top 1 & Top 5 & Top 10 & Top 15\\  
\hline 
2&0.826&0.539&0.539&0.552\\
3&0.833&0.578&0.519&0.513\\
4&0.703&0.589&0.541&0.494\\
5&0.788&0.624&0.588&0.547\\
2 \& 3&0.944&0.656&0.583&0.621\\
2 \& 3 \& 4&0.944&0.667&0.594&0.629\\
2 \& 3 \& 4 \& 5&0.944&0.667&0.622&0.658\\
\hline 
\end{tabular}
\caption{Average precision of using expansion features of different configurations of cycle lengths.} 
\label{table:cycleResults} 
\end{table}

\begin{figure*}
\centering
\begin{minipage}{.45\textwidth}
  \centering
  \includegraphics[width=.8\linewidth]{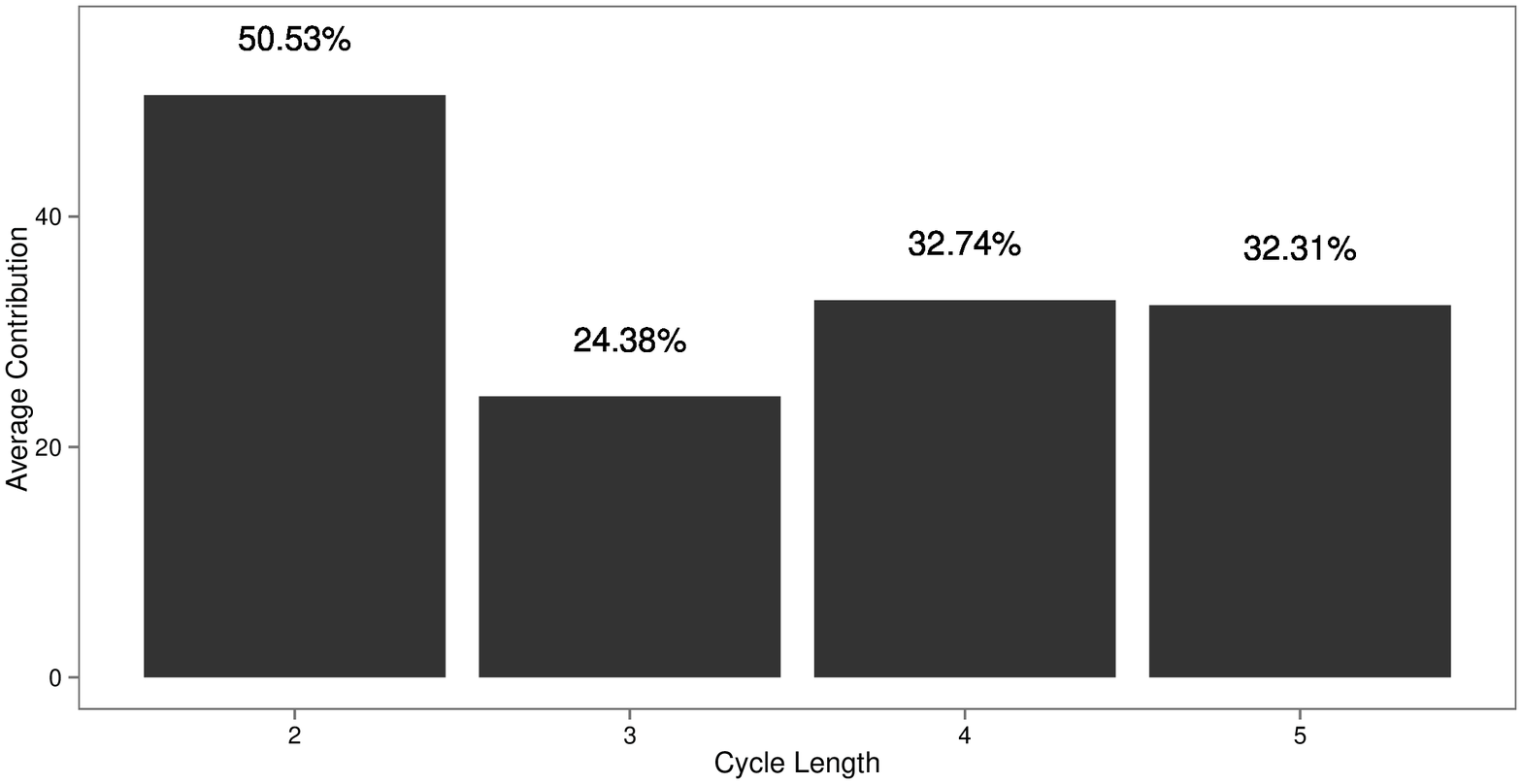}
  \captionof{figure}{Contribution vs. Cycle Length}
  \label{fig:CycleLenghtContribution}
\end{minipage}
\begin{minipage}{.45\textwidth}
  \centering
  \includegraphics[width=.8\linewidth]{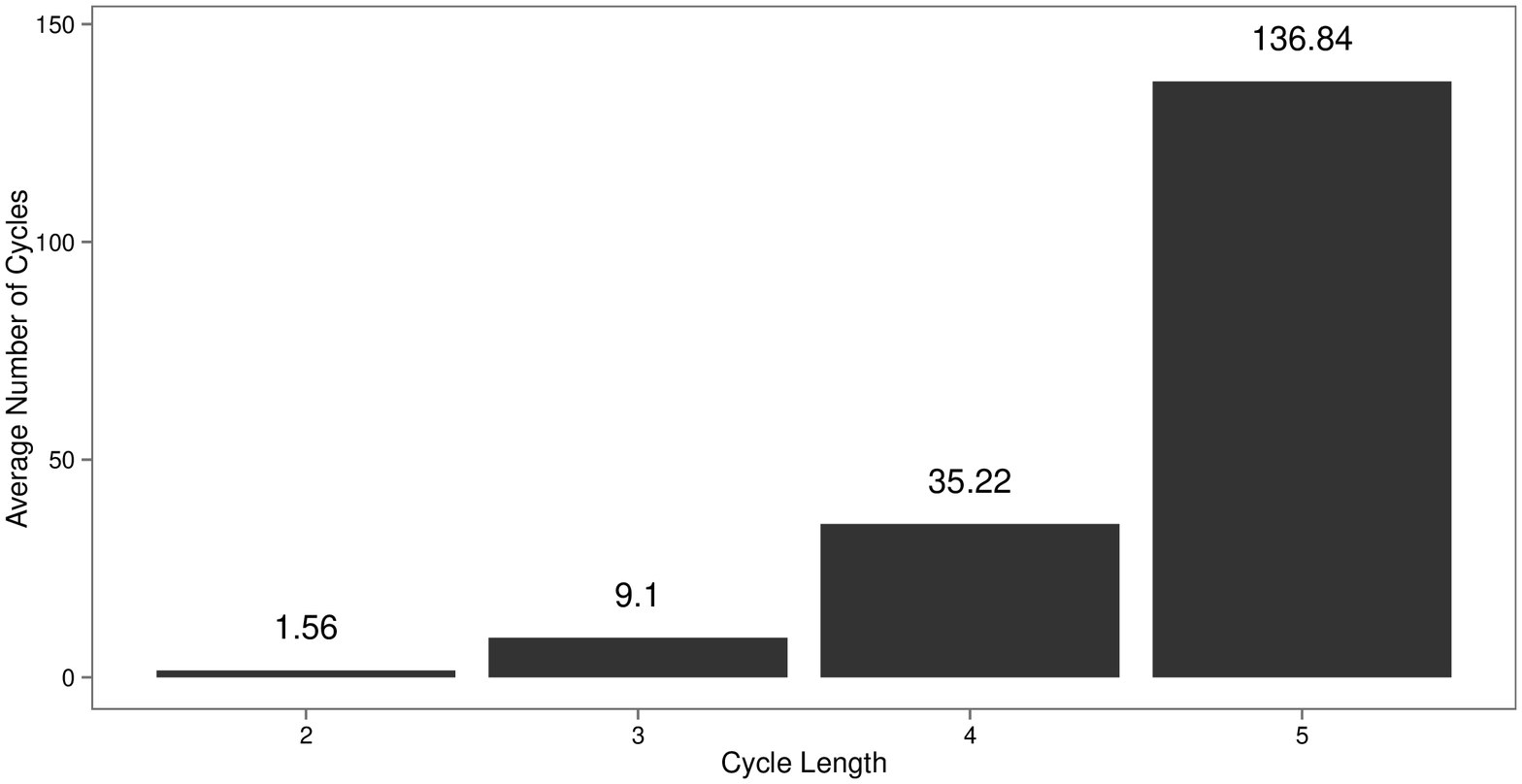}
  \captionof{figure}{Average number of cycles vs. Cycle LengthLength.}
  \label{fig:numOfCycles}
\end{minipage}%
\end{figure*}

\begin{figure*}[]
\begin{centering}
    \begin{subfigure}[b]{0.45\textwidth}
    \centering
      \includegraphics[width=.7\linewidth]{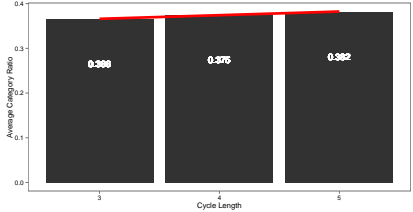}
      \caption{Category Ratio vs. Cycle Length}
      \label{subfig:categoryRatio}
    \end{subfigure}
    \begin{subfigure}[b]{0.45\textwidth}
    \centering
      \includegraphics[width=.7\linewidth]{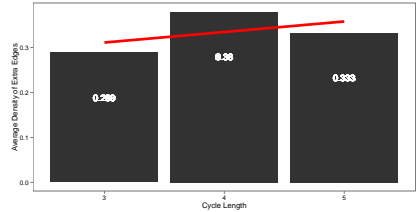}
      \caption{Density of Extra Edges vs. Cycle Length}
      \label{subfig:densityOfExtraEdges}
    \end{subfigure}
    \caption{Characteristics of cycles of length 2, 3, 4 and 5.}
    \label{fig:CyclesCharacteristics}
    \end{centering}
  \end{figure*}

In Table~\ref{table:cycleResults} we summarize the results achieved by using
the titles of the articles within the cycles of a given length, and
combinations of them, as expansion features. Broadly speaking, the precisions
achieved by the different configurations  are comparable to the best results
obtained in the ImageCLEF 2011 conference~\cite{tsikrikaPK11}.  However, the
current results of the conference were achieved by using a hybrid \mbox{--visual} and
textual-- search engine and also using relevance feedback techniques.  This
supports the idea that Wikipedia encodes relevant information in its structure
(and more concretely, in the form of cycles of articles and categories) that
can be used to solve queries from many different domains.   Also, we see that
cycles of length 2 and 3 achieve better precision on the top-1 and the top-5
results, while cycles of length 4 and 5 achieve  better results on the top-10
and top-15. This is because shorter cycles  contain articles whose title is
semantically very close and similar to $q.k$,  which are those that better
help to define the user needs. On the other hand, larger cycles contain
articles whose titles are capable  of introducing new concepts -- also
semantically related to $q.k$ -- that widen the space of search, thus, those
concepts may not be so exact but  contribute to find more results, improving
the top-10 and top-15 precision.

We define the contribution of a cycle $C$ for a query $q$ as the percentual
difference between $\mathcal{O}(\mathcal{L}(q.k),q.D)$ and \newline
$\mathcal{O}(\mathcal{L}(q.k) \cup C,q.D)$\footnote{In $\mathcal{L}(q.k) \cup C$ we
only consider the articles in $C$ but ignore the categories.}. In
Figure~\ref{fig:CycleLenghtContribution}, we show the average contribution of
cycles of different lengths. We observe that cycles of length 2 are able to
achieve an average contribution of up to 50\%, while those of in larger cycles
contribute 32.74\% at most. This suggests that cycles of length 2 contain
significantly better articles than the rest of the cycles, therefore one could
be tempted to deliberately add such cycles to expand queries. However, to
better understand these results, we also count the average number of cycles of
each length, which are shown in Figure~\ref{fig:numOfCycles}. We observe that
the amount of cycles of length 2 is significantly smaller than those larger.
This could be caused either because a) Wikipedia does not contain a large
amount of such cycles or b) because the cycles of length 2 are not always
reliable, as otherwise they would appear more frequently in the query graphs.
However, according to our experiments, among all pairs of articles that are
connected, 11.47\% form a cycle of length 2, meaning that this
structure is not so infrequent. Then, we must assume the hypothesis that the
cycles of length 2 that contribute significantly to the quality of the results
are scarce.



\begin{figure}[h!]
      \centering
      \includegraphics[width=.38\linewidth]{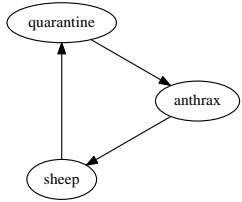}
      \caption{Category-free cycle of length 3.}
      \label{fig:wrongCycles}
      \vspace{-.1cm}
  \end{figure}

We count the ratio between categories and the total number of nodes that forms
the cycle to understand the importance of this type of nodes. Note that, due
the schema depicted in Figure~\ref{fig:wikipedia}, only cycles
whose length is equal or larger to 3
can contain categories. In
Figure~\ref{subfig:categoryRatio}, we see that among all analyzed cycles, the
average ratio of categories grows very slowly-- the slope of the trend line is
almost 0 -- when the length grows. More concretely, the number of categories
in cycles of length 3 is in general 1 ($3\cdot0.366\approx 1$), while the
number of categories in cycles of length 5 is, in general, 2 ($5\cdot 0.382
\approx 2$). This suggests that categories play a significant role in
connecting semantically related articles. This observation implies breaking
away from the idea that, the shorter the cycle and larger the proportion of
articles, the stronger the relation between them is. For example, even short cycles
of length 3 that do not contain categories, as the one depicted in
Figure~\ref{fig:wrongCycles}, may introduce semantically-distant terms as can
be ``sheep'' from ``anthrax'' that are likely to diminish the retrieval
performance of a query.

 \begin{figure}[t!]
\centering
\includegraphics[width=.7\linewidth]{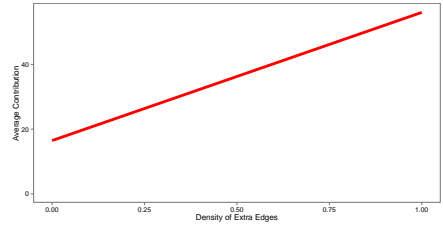}
\caption{Density of Extra Edges vs. Average Contribution.}
\label{fig:densityOfExtraEdgesTendency}
\vspace{-.5cm}
\end{figure}

Another relevant characteristic is the characterization of
the cycles based on the density of extra edges (those extra edges beside those
strictly necessary to form a cycle). The minimum amount of edges of a cycle of
length $|C|$ is $|C|$, thus we define the density of extra edges as the ratio
between the extra edges and the maximum amount of extra edges a cycle can
have. Given the following functions:\\ - $\mathcal{A}(C)$: returns the number
of articles in the cycle,\\ - $\mathcal{C}(C)$: returns the number of
categories in the cycle and\\ - $\mathcal{E}(C)$: returns the number of edges
in the cycle. \\ we calculate the maximum amount of edges cycles of length
larger than 3 as follows:

{\small
\begin{align*}
M(C) =&\ \mathcal{A}(C)\cdot(\mathcal{A}(C)-1) 
\ +&\mathcal{A}(C)\cdot\mathcal{C}(C) 
+&\frac{\mathcal{C}(C)\cdot(\mathcal{C}(C)-1)}{2}
\end{align*}
}%
and the density of extra edges is calculated as:
\begin{eqnarray*}
\frac{\mathcal{E}(C)-|C|}{M(C)-|C|}
\end{eqnarray*}

In Figure~\ref{fig:densityOfExtraEdgesTendency} we show the trend line of  the
density of extra edges compared to the contribution of the cycle.  We see
that, the denser the cycle, the better its contribution. This assertion is also
supported by the information depicted in Figure~\ref{fig:CycleLenghtContribution} and in
Figure~\ref{subfig:densityOfExtraEdges}. In particular, 
we observe that there is a correlation between the cycles that are denser in Figure~\ref{subfig:densityOfExtraEdges}
and the cycles that contribute more, depicted in Figure~\ref{fig:CycleLenghtContribution}.
Thus, cycles of length 4 are the densest and the ones that achieve 
the largest average contribution, and cycles of length 3 are the least dense and 
also the ones that achieve the smallest contribution.

\section{Conclusions and challenges}~\label{challenge}
In this paper, we have analyzed the use of knowledge bases as graphs to improve
the query expansion problem. We have created a ground truth that relates a
query with a graph of entries within Wikipedia (our knowledge base), and we
have named this the query graph. In particular, we have used the ImageCLEF
query set and Wikipedia, creating a graph of articles and categories for each
query.

Later, we have analyzed the query graphs, aiming at revealing structures that
help to extract information for the particular case of query expansion. In
other words, given a query $q$ and its query graph $G(q)$, we analyze the
structures of $G(q)$. This analysis allows us to identify the cycles as an
important structure  in order to find expansion features. According to our
analysis, dense cycles, in which the ratio of categories stands around the
30\%, are specially useful to identify new expansion features. Among the
cycles that fulfill  those properties, small cycles help to describe better
the user needs, expressed as $q.k$, and improve the precision in the first
results while, larger cycles introduce expansion features, that widen the
search space and, thus, favor the precision in the top-10 and top-15 results.

The structural analysis of the developed ground truth and its conclusions pose
some interesting challenges for graph technologies. The computation of all the
dense cycles of a given length without taking the edges direction into account
is a complex problem and computationally expensive even for a high performance
graph database. As an example, analyzing the query graphs, which have an
average size of 208.22 nodes, took us an average time of 6 minutes per query.
Taking into account that query expansion techniques are expected to respond in
real time, and that Wikipedia has almost $5M$ articles, there is still a lot
to do in many fields such as high performance graph technology and algorithms.

In the particular scenario of using Wikipedia as a knowledge base, it is  also
interesting to study the convenience of using the redirect articles  as
expansion features, since they represent less common ways to refer  a concept
and may reveal as a way of introducing good expansion featured. However, due to
the cycle analysis that we have done, redirects are never considered as an
expansion feature since they can never close a cycle (see
Figure~\ref{fig:wikipedia}).

We have not analysed how the frequency of a given article in the cycles and the
goodness of its title as expansion feature are correlated, as cycles are
considered individually. Such correlation, if existing, could be exploited.

Last but not least, future research should include techniques aimed at taking
advantage of the trends analyzed in this paper in real query expansion system,
which are expected to respond in real time.

\section*{Acknowledgments}  
{\scriptsize The members of DAMA-UPC thank the Ministry of
Science and Innovation of Spain, Generalitat de Catalunya, for grant numbers
TIN2013-47008-R and SGR2014-890 respectively and also the EU FP7/2007-2013 for
funding the LDBC project (ICT2011-8-317548). Also, thanks to Oracle Labs for
the support to our research on graph technologies.  }

\bibliographystyle{plain} 
{\scriptsize
\bibliography{grades}
}

\end{document}